\documentstyle[12pt]{article}
\def\bsh{\backslash}
\def\bdot{\dot \beta}
\def\adot{\dot \alpha}

\newfont{\bbbold}{msbm10 scaled \magstep1}

\def\bbF{\mbox{\bbbold F}}

\newfont{\goth}{eufm10 scaled \magstep1}

\def\a{\alpha}
\def\b{\beta}
\def\c{\gamma}\def\cdot{\dot\gamma}
\def\d{\delta}\def\D{\Delta}

\def\f{\phi}\def\F{\Phi}

\def\l{\lambda}

\def\p{\pi}

\def\th{\theta}

\def\beq{\begin{equation}}\def\eeq{\end{equation}}
\def\beqa{\begin{eqnarray}}\def\eeqa{\end{eqnarray}}
\def\barr{\begin{array}}\def\earr{\end{array}}

\def\del{\partial}

\def\xz{\times}
\begin{document}

\begin{titlepage}
\begin{flushright}
CERN-TH/96-159\\
King's College/kcl-th-96-13\\
hepth/9606
\end{flushright}
\vskip 2cm
\begin{center} {\bf
\Large{Operator Product Expansions in Four-dimensional Superconformal Field
Theories}}
\end{center}
\vskip 1.5cm
\centerline{\bf P.S. Howe\footnote[1]{Permanent address: Dept. of Mathematics,
King's College,
London}}
\vskip5mm
\centerline{CERN}
\centerline{Geneva, Switzerland}
\vskip 5mm
\centerline{and}
\vskip 5mm
\centerline{\bf P.C. West}
\vskip 5mm
\centerline{Department of Mathematics}
\centerline{King's College, London}
\vskip 1.5cm
\nopagebreak \begin{abstract}
\noindent
The operator product expansion in four-dimensional superconformal field theory
is discussed. It is demonstrated that the OPE takes a particularly simple form
for certain classes of operators. These are chiral operators, principally of
interest in theories with $N=1$ or $N=2$ supersymmetry, and analytic operators,
of interest in $N=2$ and $N=4$. It is argued that the Green's functions of such
operators can be determined up to constants.
\end{abstract}
\vskip 3cm
\begin{flushleft}
CERN-TH/96-159\\
\today
\end{flushleft}
\end{titlepage}

A central r\^{o}le in two-dimensional conformal field theories is played by
operator product expansions \cite{bpz}. Indeed, all the properties of  these
theories
can be encoded in their OPE's. The OPE of two primary fields yields not
only primary fields, but also descendants of primary fields. However, strong
constraints on the OPE follow from demanding that it be compatible with
conformal
symmetry. This procedure determines
the space-time dependent coeficients with which the primary fields occur up
to constants of proportionality. These constants, however,  can only be
determined by the details
of the model, in effect what  null states it possess. Given these constants,
the
rest of the OPE, that is all the dependence on the descendants, is
determined
by conformal symmetry. In particular, in minimal models, the OPE's close with
only a finite number of primary fields, so that the correlation functions of
such theories can in principle be calculated using the OPE and the two and
three-point functions.

In four dimensions conformal field theories were studied in some detail in a
general setting some time ago \cite{tmp}, but at that time no non-trivial
examples were known. However, it is now known that there are many
supersymmetric gauge theories which are superconformal, certainly in
perturbation theory and perhaps beyond. These theories are: the maximally
supersymmetric ($N=4$) models \cite{sw}, a class of $N=2$ \cite{hsw} models and
certain $N=1$ theories \cite{pw}. It has further been conjectured that more
supersymmetric theories may have non-trivial fixed points \cite{w}. It is
therefore appropriate to reconsider four-dimensional operator product
expansions in the supersymmetric context.

It has been observed that the chiral sector of superconformally invariant
theories in
four dimensions has certain similarities with such sectors  in two dimensional
superconformal theories. In
particular,  the chiral and dilation weights of chiral operators are
related if the theory is at a fixed point \cite{fiz}. It has also been pointed
out that the so-called analytic sectors of $N=2$ and $N=4$ theories seem to
have similar properties \cite{hw}.
Moreover, it has been argued that, although these supersymmetric theories
are only invariant under a finite dimensional superconformal group, their
very special form allows one to solve, non-perturbatively, for large classes
of their Green's functions. In particular, one can determine the Green's
functions in any chiral or anti-chiral  sector and it is likely that one can
also do this in the analytic sector \cite{hw}. In this paper, we
give the operator product expansions in these sectors and find close
similarities to the corresponding two dimensional results.

We begin by giving a discussion of operator product expansions and their
conformal properties which applies in a general setting. For simplicity we
shall take spacetime or superspace to be complex. Let us denote
the complex, finite-dimensional (super)conformal group by $G$,  for example,
for four-dimensional $N$-extended
supersymmetry $G=SL(4|N)$. The (super)conformal theories of interest to us
are described by (super)fields
which live on the  (super)space $P\bsh G$
where $P$ is a parabolic subgroup of $G$. Which subgroups one should take in
four dimensions can be found in
reference \cite{hh,hw}. We will denote the  coordinates of  $P\bsh G$
by $X$.

We define primary fields to be fields that transform under an induced
representation of $G$. To keep life simple, we shall suppose that the fields
are one-dimensional, i.e. transform under a one-dimensional subgroup of $P$. In
many cases such fields are the most interesting to consider.
For an infinitesimal (super)conformal transformation we have
\beq
\d\f=V\f + q\D\f
\eeq
where $q$ is a charge associated with the representation (related to the
dilation weight of the field) and $V$ is the vector field generating the
transformation on the coset space,
$V(X)=\delta X{\partial\over{\partial X}}$,
$\delta  X$ being the change in $X$. The function $\D$ is a function that
characterises the induced representation. We can choose coordinates such that
the components of $V$ are polynomials of degree 2 in the components of $X$ and
such that $\D$ is a polynomial of degree 1.
Descendants
are space-time or superspace derivatives of primary fields. These will not in
general transform as induced representations. Indeed, under certain conformal
transformations descendants mix into primary fields. One can also take
descendants to be given by group generators acting on the primary fields, but
this is equivalent to the above desciption.

Now consider a complete set of operators $\{\F_I\}$ comprising both primary
fields $\{\f_i\}$ and their descendants. We shall assume that we can write an
operator product expansion in the standard form,
\beq
\F_I(X_1) \F_J(X_2)=\sum_K f_{IJ}^K(X_1,X_2) \F_K(X_2)\ .
\eeq
Applying an infinitesimal (super)conformal transformation to this and
considering only primary fields on the left-hand side we find
\beq
\sum_K [(V_1+V_2 +q_i\D_1 +q_j\D_2)f_{ij}^K]\F_K=\sum_K
f_{ij}^K(\d\F_K(2)-V_2\F_K(2))\ ,
\eeq
where the subscripts 1 and 2 refer to the two points involved. Under a
transformation for which $\D$ is $X$-independent, the terms proportional to the
primary fields do not have any contributions from the transformations of the
descendants and we get
\beq
(V_1+V_2 +(q_i +q_j-q_k)\D)f_{ij}^k(1,2)=0\ .
\eeq
Hence, for these transformations, the coefficients $f_{ij}^k$ behave as a two
point function with a total  $q$ weight of  $(q_i +q_j-q_k)$. For the ordinary
conformal
group these transformations are
translations, dilations and Lorentz rotations. However, these transformations
are sufficient
to determine $f_{ij}^k$ up to a constant. In particular, if the fields under
consideration are Lorentz scalars and the primary fields have dilation weights
$d_i(=q_i$ in this case) then
\beq
f_{ij}^k={{c_{ij}^k}\over{[(x_1-x_2)^2]^{{1\over2}(d_i+d_j-d_k)}}}\ ,
\eeq
where $c_{ij}^k$ are constants and $x_1$ and $x_2$ are  the positions
of the
primary fields $\f_i$ and $\f_j$ in spacetime.
\par
For the remaining transformations the function  $\D$
is (super)
space dependent. However, we can
still examine only the primary field terms provided we take into account of
transformations
of descendants which result in primary fields. This calculation determines
the coefficients of the lowest level descendants.

Let us illustrate the
procedure for the simplest situation, i.e. the ordinary conformal group in four
dimensions. The only
remaining
transformations are the special conformal transformations with parameter
$C_{\dot\beta\beta}$ for which
$\D=x^{\beta\dot\beta}C_{\dot\beta\beta}$.
Under this transformation only the lowest descendants, i.e. the set of fields
$\{\partial_{\alpha\dot\alpha}
\f_i\}$, transform into primary fields. Including these terms explicitly,
the OPE of
equation (3) becomes
\beq
\f_i(x_1)\f_j(x_2)=\sum_k f_{ij}^k(x_1,x_2)\f_k(x_2)
+\sum_k
f_{ij}^{k;\alpha\dot\alpha}(x_1, x_2)
\partial_{\alpha\dot\alpha}\f_k (x_2)
+\ldots
\eeq
where the dots denote contributions from higher order descendants.
Applying our previous argument, again except for special conformal
transformations, we recover (5) and find that
\beq
f_{ij}^{k;\alpha\dot\alpha} (x_1,
x_2)={{(x_{12})^{\alpha\dot\alpha}}\over{(x_{12}^2)
^{{1\over2}(d_i+d_j-d_k)}}} c_{ij}^{k(1)}\ ,
\eeq
 where the $c_{ij}^{ k (1)}$ are constants and $x^{\alpha\dot\alpha}_{12}=
x_1^{\alpha\dot\alpha}-x_2^{\alpha\dot\alpha}$. Applying a special conformal
transformation we find
\beq
c_{ij}^{k(1)}=c^k_{ij}{{(d_i-d_j+d_k)}\over{2d_k}}\ .
\eeq

By carrying out all conformal transformations on each side of the
OPE and
comparing coefficients of the descendant fields we can determine all the
descendant contributions
in terms of the constants $c_{ij}^k$. Thus the situation is the same as in two
dimensional conformal field theories.

We now apply the above procedure to  chiral superfields in four-dimensional
$N$-extended
supersymmetry. The analysis can be adapted straightforwardly to other
dimensions where chiral fields are available. Due to the chiral constraint,
a chiral superfield can be viewed as a function of only $x^{\alpha \dot\alpha}$
and
$\theta
^{\alpha a}$, where $x$ is an appropriate chiral variable and $a=1,\ldots N$.
The operator product expansion of two chiral superfields
$\f_i(X_1)$ and $\f_j(X_2)$ can be written as
\beq
\barr{lcl}
\f_i(X_1)\f_j(X_2)
&=&\sum_k\{f_{ij}^k (X_1,X_2) \f_k
(X_2)+f_{ij}^{k;\alpha a }  (X_1,X_2) (\del_{\a a}\f_k)(X_2)+\\
& &+f_{ij}^{k,\alpha\dot\alpha}  (X_1,X_2)
(\partial_{\alpha\dot\alpha}\f_k)(X_2)+
\ldots\}\ ,
\earr
\eeq
where again the dots denote contributions from higher order descendants and
where we have used the shorthand notation
$\partial_{\alpha \dot \alpha}={\partial \over \partial x^{\alpha
\dot \alpha}}$ and
$\partial _{ \alpha a}= {\partial \over \partial \theta
^{\alpha  a}}$.

We now give the  superconformal transformations written in terms of the
chiral variables.
The
vector fields which generate the
translations $(P)$, dilations $(D)$ and  special conformal
transformations $(K)$ are:
\beq
\barr{lcl}
V(P)_{\a\adot}&=&\del_{\a\adot}\\
V(D)&=&x^{\a\adot}\del_{\a\adot}\\
V(K)^{\a\adot}&=& x^{\a\bdot}x^{\b\adot}\del_{\b\bdot}+x^{\b\adot}\th^{\a
a}\del_{\b a}\ ,
\earr
\eeq
and the associated $\D$'s are
\beq
\barr{lcl}
\D(P)_{\a\adot}&=& 0\\
\D(D)&=& 1 \\
\D(K)^{\a\adot}&=& x^{\a\adot}\ .
\earr
\eeq
The vector field generating internal symmetry $(I)$ transformations ($SL(N)$)
is
\beq
V(I)_a{}^b=\th^{\a b}\del_{\a a}-{1\over N}\d_a^b \th^{\a
c}\del_{\a c}
\eeq
and the function $\D(I)_a^b$ vanishes in this case. For $N\neq 4$ we also have
$R$-symmetry transformations generated by
\beq
V(R)=\th^{\a a}\del_{\a a}
\eeq
with
\beq
\D(R)={2N\over (N-4)}\ .
\eeq
The $Q$-supersymmetry transformations are generated by
\beq
\barr{lcl}
V(Q)_{\a a}&=& \del_{\a a}\\
V(Q)_{\adot}^a &=& -\th^{\a a}\del_{\a\adot}\ ,
\earr
\eeq
and the $S$-supersymmetry generators are
\beq
\barr{lcl}
V(S)_a^{\adot} &=& x^{\a\adot}\del_{\a a} \\
V(S)^{\a a}&=& -x^{\a\bdot}\th^{\b a}\del_{\b\bdot} +\th^{\a b}\th^{\b
a}\del_{\b b}\ ,

\earr
\eeq
and only the last of these has a non-vanishing $\D$ given by
\beq
\D(S)^{\a a}=-\th^{\a a}\ .
\eeq
There are also Lorentz transformations which act in the obvious way on the
vector and spinor coordinates.

The transformations for which $\D$ is constant can, according to our general
arguments, be used to determine the superspace
dependence of the coefficients of the primary chiral superfields.
Translations and supersymmetry transformations imply that the coefficients are
functions of $x_{12}^{\alpha\dot\alpha}\equiv x_1^{\alpha\dot\alpha}-x_2
^{\alpha\dot\alpha}$ and $\theta_{12}^{\alpha a}=\theta_1^{\alpha a}-\theta_2
^{\alpha a}$. $R$ symmetry implies that if $f_{ij}^k$  is to be non-zero
it must be proportional to $\theta_{12}^{\alpha a}$ to the power $q_i+q_j-q_k$.
(There are no chiral fields of interest in $N=4$ rigid supersymmetry, so this
is always valid for the applications we have in mind.) Let
us consider in detail the case when $q_i+q_j=q_k$. Dilation and Lorentz
symmetry imply that
\beq
f_{ij}^k=c_{ij}^k,\quad f_{ij}^{k;\alpha a}=\theta^{\alpha a}
c_{ij}^{k(2)}
\eeq
and
\beq
f_{ij}^{k,\alpha\dot\alpha}=(x_{12})^{\alpha\dot\alpha}c_{ij}^{k(3)}\ .
\eeq
To fix the descendant coefficients we use special conformal transformations and
special ($S$) supersymmetries. We find
that the contribution given by $\f_k$ and its descendants to the OPE is
\beq
\barr{lcl}
\f_i(X_1)\f_j(X_2)&=& c_{ij}^k\left\{\f_k(X_2)+
{{q_i}\over{q_k}}\theta_{12}^{\alpha a}(\del_{\a a}
\f_k)(X_2)
+{{q_i}\over{q_k}}x_{12}^{\alpha\dot\alpha}(\partial_{\alpha\dot\alpha}
\f_k) (X_2)\right\}\\
& & + {\rm {higher\ order\ descendants}}\ .
\earr
\eeq
This result is essentially identical to the analogous result for two
dimensional
superconformal field theory.

We may also have contributions from primaries which are Lorentz scalars and
which have $q_i+q_j-q_k=3$, for $N=1$, and $q_i+q_j-q_k=2$, for $N=2$. Such
terms have leading contributions of the form
\beq
f_{ij}^k=c_{ij}^k {\th_{12}^2\over x_{12}^4}
\eeq
for $N=1$, and
\beq
f_{ij}^k=c_{ij}^k {\th_{12}^4\over x_{12}^4}
\eeq
for $N=2$. There are also primary fields with undotted spinor indices and
internal indices. For example, in $N=1$, one can have a contribution to the OPE
of Lorentz scalars $\f_i$ and $\f_j$ from a spin one-half field $\f_{k\a}$ with
charge $q_k$ if $q_i+q_j-q_k={3\over2}$, and for which the leading contribution
would be
\beq
f_{ij}^{k\a}=c_{ij}^k{\th_{12}^{\a}\over x_{12}^2}
\eeq
However, for any pair of primary chiral fields, one always finds a finite
number of primaries in the OPE on the right-hand side determined by the charges
and spinorial representations involved.

We now consider harmonic superfields \cite{gikos}. For the theories of most
interest
to us, i.e. the extended rigidly supersymmetric theories, superfields of this
type occur
in $N=4$ Yang-Mills theory and in the $N=2$ matter sector of $N=2$ theories. To
be concrete we consider
the former case but the formalism can be easily adapted to $N=2$. The $N=4$
harmonic superspace of interest to us is the extension of Minkowski superspace
by the internal space $\bbF=S(U(2)\times U(2))\bsh SU(4)$, and the fields we
wish to consider are analytic fields on this space, that is to say, fields
which are analytic with respect to the internal space $\bbF$, and which are
also Grassmann analytic ($G$-analytic). The latter means that they are
annihilated by half of the superspace covariant derivatives, and therefore
depend on only half of the odd coordinates, in a similar fashion to chiral
fields. The difference is that the derivatives involve the coordinates of the
internal space and this allows one to use a mixture of dotted and undotted
spinor derivatives.
These
fields can be defined on a new superspace, analytic superspace, which is
similar to chiral superspace. It has local coordinates
\[
X=\{x^{\alpha\dot\alpha},\ \lambda^{\alpha a^\prime},\
\pi^{a\dot\alpha},
\ y^{aa^\prime}\}
\]
where $a$ and $a'$ can both take on two values. (Locally, the internal space is
just like ordinary complex Minkowski space).

The operator product expansion for two analytic fields takes the form
\beq
\barr{lcl}
\f_i(X_1)\f_j(X_2)&=&\sum_k\{f_{ij}^k (X_1, X_2)
\f_k(X_2)
+f_{ij}^{k,\alpha\dot\alpha} (X_1, X_2)
(\del_{\a\adot}
\f_k)(X_2)\\
&&+f_{ij}^{k,a\dot\alpha} (X_1, X_2)
(\del_{a\adot}\f_k)(X_2)
+f_{ij}^{k,\alpha a\prime} (X_1, X_2)
(\del_{\a a'}
\f_k)(X_2)\\
&&+f_{ij}^{k,aa\prime} (X_1, X_2)
(\del_{aa'}\f_k)
(X_2)+\ {\rm{higher\ order\ descendants}}\}\ .
\earr
\eeq

The superconformal transformations, when written in  analytic
coordinates, take a particularly simple form.
The vector fields which generate them are, for translations, dilations, Lorentz
transformations ($M$) and special conformal transformations,
\beq
\barr{lcl}
V(P)_{\a\adot} &=& \del_{\a\adot} \\
V(D)           &=& x^{\a\adot}\del_{\a\adot} +{1\over2}\l^{\a a'}\del_{\a
a'}+{1\over2} \p^{a\adot}\del_{a\adot} \\
V(M)_{\a}^{\b} &=&(x^{\a\cdot}\del_{\a\cdot}+\l^{\a c'}\del_{\a c'})-{\rm
trace}\\
V(M)_{\adot}^{\bdot}
&=&(x^{\c\bdot}\del_{\c\adot}+\p^{c\bdot}\del_{c\adot})-{\rm trace}\\
V(K)^{\a\adot} &=& x^{\a\bdot} x^{\b\adot}\del_{\b\bdot}+x^{\b\adot}\l^{\a
b'}\del_{\b b'}+\p^{b\adot} x^{\a\bdot}\del_{b\bdot}+\p^{b\adot}\l^{\a
b'}\del_{b b'}\ .
\earr
\eeq
For internal symmetry transformations they are
\beq
\barr{lcl}
V(I)_{a a'}    &=&\del_{a a'} \\
V(I)           &=&y^{a a'}\del_{aa'}+{1\over2}\l^{\a a'}\del_{\a a'}
+{1\over2}\p^{a\adot}\del_{a\adot} \\
V(I)_a^b       &=&(\p^{b\cdot}\del_{a\cdot}+y^{bc'}\del_{ac'})-{\rm trace}\\
V(I)_{a'}^{b'} &=&(\l^{\c b'}\del_{\c a'}+y^{cb'}\del_{ca'})-{\rm trace}\\
V(I)^{aa'}     &=&y^{ab'}y^{ba'}\del_{bb'}+\l^{\b a'}y^{a b'}\del_{\b b'}+y^{b
a'}\p^{a\bdot}\del_{b\bdot}+\l^{\a a'}\p^{a\bdot}\del_{\b\bdot}\ .
\earr
\eeq
For $Q$-supersymmetry transformations we have
\beq
\barr{lcl}
V(Q)_{\a a'}     &=& \del_{\a a'} \\
V(Q)_{a\adot}    &=&\del_{a\adot} \\
V(Q)_{\a}^{a}    &=&y^{a b'}\del_{\a b'}+\p^{a\adot}\del_{\a\adot} \\
V(Q)_{\adot}^{a'}&=&y^{ba'}\del_{b\adot}-\l^{\b a'}\del_{\b\adot}\ ,
\hspace{2.3in}
\earr
\eeq
while for $S$-supersymmetry we have
\beq
\barr{lcl}
V(S)_a^{\a}    &=& x ^{\a \bdot}\partial _{a
\dot\beta}  +\lambda ^{\a b'}\partial_{a b'}
\\
V(S)_{a'}^{\adot}  &=& x ^{\b
\adot }\partial_{\b a^\prime } - \pi^{b \adot}\partial_{b a^\prime}
 \\
V(S)^{\adot a} &=&   x^{\b \adot} y^{a b'}\partial_{\b b' } - \pi^{b\adot} y^{a
b^\prime}\partial_{b b^\prime}  +
 x ^{\b \adot}\pi^{a\bdot}\partial_{\b \dot\beta}
 - \pi^{b \adot} \pi^{a\bdot}\partial _{b \bdot} \\
V(S)^{\a a'} & = &  y^{b a ^\prime} x^{\a \dot\beta }\partial _{b\bdot}  + y^{b
a ^\prime} \lambda ^{\a b^\prime}\partial_{b b^\prime}  -
\lambda ^{\b a^\prime} x^{\a\bdot }\partial_{\b\bdot}
-\lambda ^{\b a^\prime} \lambda ^{\a
b^\prime}\partial_{\b b^\prime }\ .
\earr
\eeq
The non-zero $\D$'s are
\beq
\barr{lcl}
\D(K)^{\a\adot}   &=&x^{\a\adot}\\
\D(I)             &=& -1 \\
\D(I)^{a a'}      &=&-y^{a a'} \\
\D(S)^{a\adot}    &=&\p^{a\adot} \\
\D(S)^{\a a'}     &=&-\l^{\a a'}\ .
\earr
\eeq
The transformation for an analytic field with charge $q$ takes the form given
in equation (1); the charge is the charge of the field with respect to the
internal $U(1)$, i.e. the $U(1)$ of the isoptropy group of the internal space
$\bbF$. In harmonic superspace it would correspond to a field satisfying
$D_o\f=q\f$ where $D_o$ is the derivative on $SU(4)$ corresponding to this
$U(1)$. In the case of $N=2$ there is also an $R$-symmetry transformation
generated by
\beq
V(R)=\l^{\a}\del_{\a}-\p^{\adot}\del_{\adot}\ .
\eeq
(In $N=2$ analytic space the internal indices $a$ and $a'$ only take one value
and so can be dropped.)

We shall now analyse the OPE for analytic fields using the method outlined
above. One finds again that the primary coefficients $\{f_{ij}^k\}$ must obey
the same equations as a two-point function with total charge
${1\over2}(q_i+q_j-q_k)$, i.e.,
\beq
(V_1 + V_2 +{1\over2}(q_i+q_j-q_k)(\D_1 + \D_2))f_{ij}^k=0\ .
\eeq
Now the basic two-point function in $N=4$ is the one for an Abelian Yang-Mills
field strength tensor $W$ which has charge $q=1$. It is given by
\beq
<W(X_1)W(X_2)>\propto{\hat y^2\over x^2}:=g_{12}
\eeq
where
\beq
\hat y^{a a^\prime }= y^{a a^\prime }
+2{ \lambda^{\alpha a^\prime}\pi^{a \dot \alpha}  x_{\alpha \dot
\alpha}\over x^2}\ .
\eeq
One then has
\beq
f_{ij}^k=c_{ij}^k (g_{12})^{{1\over2}(q_i+q_j-q_k)}\ .
\eeq
for some constants $\{c_{ij}^k\}$.

We can determine the coefficients for the descendant fields in the same way as
before and so we arrive at the result
\beq
\barr{lcl}
\f_i(X_1)\f_j(X_2)&=&\sum_{k=0}c_{ij}^k(g_{12})^{{1\over2}(q_i+q_j-q_k)}
\{\f_k(X_2)+{1\over2}(q_i-q_j+q_k)\xz \\
& &\xz (x_{12}^{\a\adot}\del_{\a\adot}+\l_{12}^{\a a'}
\del_{\a a'} +\p_{12}^{a\adot}\del_{a\adot}+y^{a a'}_{12}\del_{ a
a'})\f_k(X_2)\} +\ldots\ .
\earr
\eeq

In an $N=4$ Yang-Mills theory with gauge group $SU(M)$, for example, the basic
local analytic operators are given by the gauge-invariant powers of the field,
i.e., they are the operators
\beq
A_m:={\rm tr} (W)^m,\ m=2,\ldots (M-1)\ .
\eeq
The operator $A_m$ has charge $q=m$. In particular, $A_2$ is the supercurrent
which we shall denote by $T$. Its components include the energy-momentum
tensor, the spacetime supersymmetry currents and the currents corresponding to
the internal $SU(4)$ symmetry group. The OPE for two $T$'s is
\beq
\barr{lcl}
T(1)T(2)&=&c_o(g_{12})^2 +c_2g_{12}\{T(2)+(x^{\a\adot}\del_{\a\adot}+\l^{\a
a'}\del_{\a a'} +\p^{a\adot}\del_{a\adot}+y^{aa'}\del_{aa'})T(2)\}\\
&&+ \ {\rm finite\ terms}\ .
\earr
\eeq
For most four-dimensional theories the OPE of the energy-momentum tensor with
itself does not close on itself. However, for $N=4$ Yang-Mills it does, and the
result is strikingly similar to the two-dimensional case; indeed, we can
rescale $T$ such that $c_2=1$ in which case it would be tempting to interpret
$c_o$ as a
the central charge. We remark that it is only in $N=4$ that this can happen
because in $N=1$ and $N=2$ the supercurrent is neither chiral nor analytic, and
we believe that it is only these special types of superfields which have such
simple OPE's. For a discussion of the OPE in $N=1$ supersymmetric theories we
refer the reader to \cite{gj}.

There may be operators other than the $\{A_m\}$ and their descendants appearing
in the analytic OPE. For example, given a gauge-invariant scalar superfield on
super Minkowski space one can always construct a gauge-invariant analytic field
on harmonic superspace by applying enough spinorial derivatives. A similar sort
of situation can arise with chiral fields in $N=1$. If $\f$ is chiral then so
is $\bar D^2\bar\f$. However, the latter is not in general primary unless $\f$
has weight ${1\over3}$. Clearly, analytic operators obtained in this way will
only be able contribute to the analytic OPE if they are primary. The
lowest-dimensional analytic operator of this type that one can construct in
$N=4$ has na\"{\i}ve dimension 6 so that, even if it is primary, it cannot
contribute to the OPE of two supercurrents.

We conclude with a consequence of the OPE for analytic fields in either $N=2$
or $N=4$. From a formal point of view the spacetime coordinate $x$ and the
internal coordinate $y$ appear in a very symmetrical manner. Indeed, as we have
remarked earlier, in $N=4$ the internal space is locally the same as (complex)
Minkowski space. However, from a physical point of view spacetime and the
internal space are completely different. In particular, the singularities which
appear in the OPE as $x_1$ approaches $x_2$ are due to the usual difficulties
encountered in defining local products of operators in quantum field theory. On
the other hand, the r\^{o}le of $y$ is simply to act as a device to help us
exploit the internal symmetries of the theory. The internal space is compact,
and no internal points need or should be removed from the domain of definition
of Green's functions of many operators. Therefore singularities in the internal
variables are completely spurious and must cancel. One way of seeing this is to
note that any analytic operator can be reexpressed in terms of a polynomial in
$y$ with coefficients which are fields on ordinary super Minkowski space. If
one examines the right-hand side of the analytic OPE above, one sees that the
absence of singularities  in $y$ requires that ${1\over2}(q_i+q_j-q_k)$ be an
integer, and furthermore that there can only be a finite number of primary
fields occurring because otherwise one will introduce poles in $y$ for
sufficiently large values of $q_k$. Thus the situation is similar in some
respects to that obtaining in two-dimensional minimal models. Given that the
analytic OPE is valid, and that analyticity imposes finiteness of the number of
primaries occurring in any given OPE, it is tempting to conclude that any
Green's function of analytic operators can in principle be computed knowing the
three-point functions and the OPE. Any such Green's function depends only on a
few arbitrary constants, i.e. the $\{c_{ij}^k\}'s$ and the constants in the
three-point functions. In other words, the analytic OPE for $N=4$ (and for
$N=2$) suggests that this sector of these theories is solvable in the full
quantum theory. We note, however, that this result depends on some assumptions,
principally the form of the OPE for analytic fields and the assumption that
analyticity is maintained in the quantum theory. The latter seems to be natural
given that the theories we are interested in are superconformal. In a future
paper \cite{hw2} we shall give a more detailed discussion of the Green's
functions using analyticity and superconformal invariants.

\end{document}